\documentclass[preprint2]{aastex}

\newcommand{\ms}{\mbox{m s$^{-1}$}}

\shorttitle{CHIRON}
\shortauthors{Tokovinin et al.}

\begin{document}


\title{CHIRON -- A Fiber Fed Spectrometer for Precise Radial Velocities}

\author{Andrei Tokovinin\altaffilmark{1},
Debra A. Fischer\altaffilmark{2},
Marco Bonati\altaffilmark{1},
Matthew J. Giguere\altaffilmark{2},
Peter Moore\altaffilmark{1},
Christian Schwab\altaffilmark{2,3},
Julien F.P. Spronck\altaffilmark{2},
Andrew Szymkowiak\altaffilmark{2}
}

\email{atokovinin@ctio.noao.edu}

\altaffiltext{1}{Cerro Tololo Inter-American Observatory, Casilla 603, La
  Serena, Chile}

\altaffiltext{2}{Department of Astronomy, 
Yale University, New Haven, CT 06511, USA}

\altaffiltext{3}{Sagan Fellow}

\begin{abstract}
The   CHIRON   optical   high-resolution  echelle   spectrometer   was
commissioned at the  1.5\,m telescope at CTIO in  2011. The instrument
was  designed for  high throughput  and  stability, with  the goal  of
monitoring radial  velocities of bright stars with  high precision and
high  cadence  for the  discovery  of  low-mass exoplanets.   Spectral
resolution of $R=79\,000$ is attained when using a slicer with a total
(including telescope and detector)  efficiency of 6\% or higher, while
a resolution  of $R=136\,000$ is  available for bright stars.  A fixed
spectral range of  415 to 880\,nm is covered.   The echelle grating is
housed  in  a  vacuum  enclosure  and the  instrument  temperature  is
stabilized to $\pm 0.2^\circ$.   Stable illumination is provided by an
octagonal multimode fiber  with excellent light-scrambling properties.
An iodine  cell is used  for wavelength calibration.  We  describe the
main optics,  fiber feed, detector, exposure-meter,  and other aspects
of  the  instrument, as  well  as  the  observing procedure  and  data
reduction.
\end{abstract}




\section{Introduction}
\label{sec:intro}


While   increasingly  large   astronomical  telescopes   with  complex
instrumentation  are  now  being  constructed, small  and  medium-size
telescopes are still  valuable, particularly for dedicated time-domain
projects.   Monitoring  radial  velocities  (RVs)  of  stars  for  the
discovery  and characterization of  exoplanets, binary  companions, or
pulsations are a few examples of projects that can be carried out with
moderate  aperture  telescopes  equipped with  stable  high-resolution
optical  spectrographs. Recently,  several echelle  spectrometers have
been built and put  into operation, e.g. SOPHIE \citep{SOPHIE}, HERMES
\citep{HERMES},  PFS  \citep{PFS}.  This  paper  describes the  CHIRON
optical  echelle   spectrometer  at  the  1.5-m   telescope  in  Chile
\citep{CHI-2010,CHI-2012}.

The 1.5-m  telescope at Cerro Tololo  Interamerican Observatory (CTIO)
is  operated  by the  SMARTS  (Small  and  Moderate Aperture  Research
Telescope  System) consortium  where  observing time  is purchased  by
members.   In addition, 25\%  of the  time is  distributed in  an open
competition by National Optical  Astronomy Observatory (NOAO) and 10\%
is  available to Chilean  astronomers.  The  observations are  done in
service mode by telescope operators.  This is an ideal arrangement for
high-cadence   observing  programs   and   long-term  RV   monitoring.
Unfortunately, the  1.5-m is too  old for a cost  effective conversion
into a robotic facility.

In the 1980's, the {\it Bench Mounted Echelle} (BME) was developed for
the 1.5-m telescope, and light was coupled to the instrument through a
fiber \citep{Barden}. This instrument  was decommissioned in 2001, and
after the  retirement of  the echelle at  the Blanco 4-m  telescope in
2004, the observatory was left without a high-resolution spectrograph,
while FEROS  and HARPS were intensively  used at ESO, and  the Du Pont
echelle was  used at Las Campanas. In  2008, a new fiber  feed for the
CTIO  1.5-m was  constructed and  adapted to  the old  Blanco Echelle.
This  combination, called  the  {\it Fiber  Echelle},  was offered  to
SMARTS and NOAO  users from 2008 to 2010 \citep[e.g. ][]{Richardson2011},
until it was replaced by CHIRON\footnote{NSF ARRA MRI grant 0923441}.

The design for CHIRON was primarily  driven by the need to measure RVs
with high  precision for searches  for low-mass planets  around nearby
bright  stars  such  as  $\alpha$~Cen \citep{Guedes08}.   Our  funding
envelope ($\sim$\$600K) did not allow  us to put the entire instrument
in  a  vacuum enclosure  (in  which  case  a thorium-argon  wavelength
calibration could be used), so  an iodine cell was used for wavelength
calibration.   Iodine  cells  are  often used  with  traditional  slit
spectrographs and yield high RV precision even for instruments without
environmental  stabilization   \citep[e.g.][]{Fischer12}  because  the
iodine spectrum  is embedded in  the stellar spectrum and  follows the
same  optical  path  through   the  instrument. 

We considered  design trade offs between high  spectral resolution and
the  signal to  noise ratio  (SNR) that  can be  reached with  a 1.5-m
telescope.  An  important consideration is  that we limit  the maximum
exposure time for precise Doppler  measurements to 15 minutes in order
to  limit  errors in  our  barycentric  correction.   Even if  nightly
velocities  are binned  to increase  the effective  SNR, we  model the
Doppler shift in  each observation individually and a  SNR of at least
50 is  needed to  minimize errors  in our model  of the  spectral line
spread function (SLSF).  The SLSF is sometimes called the point spread
function  (PSF), however  this is  not an  accurate term  to  use when
referring to the instrumental smearing of spectral lines.

\begin{figure*}
\epsscale{1.9}
\plotone{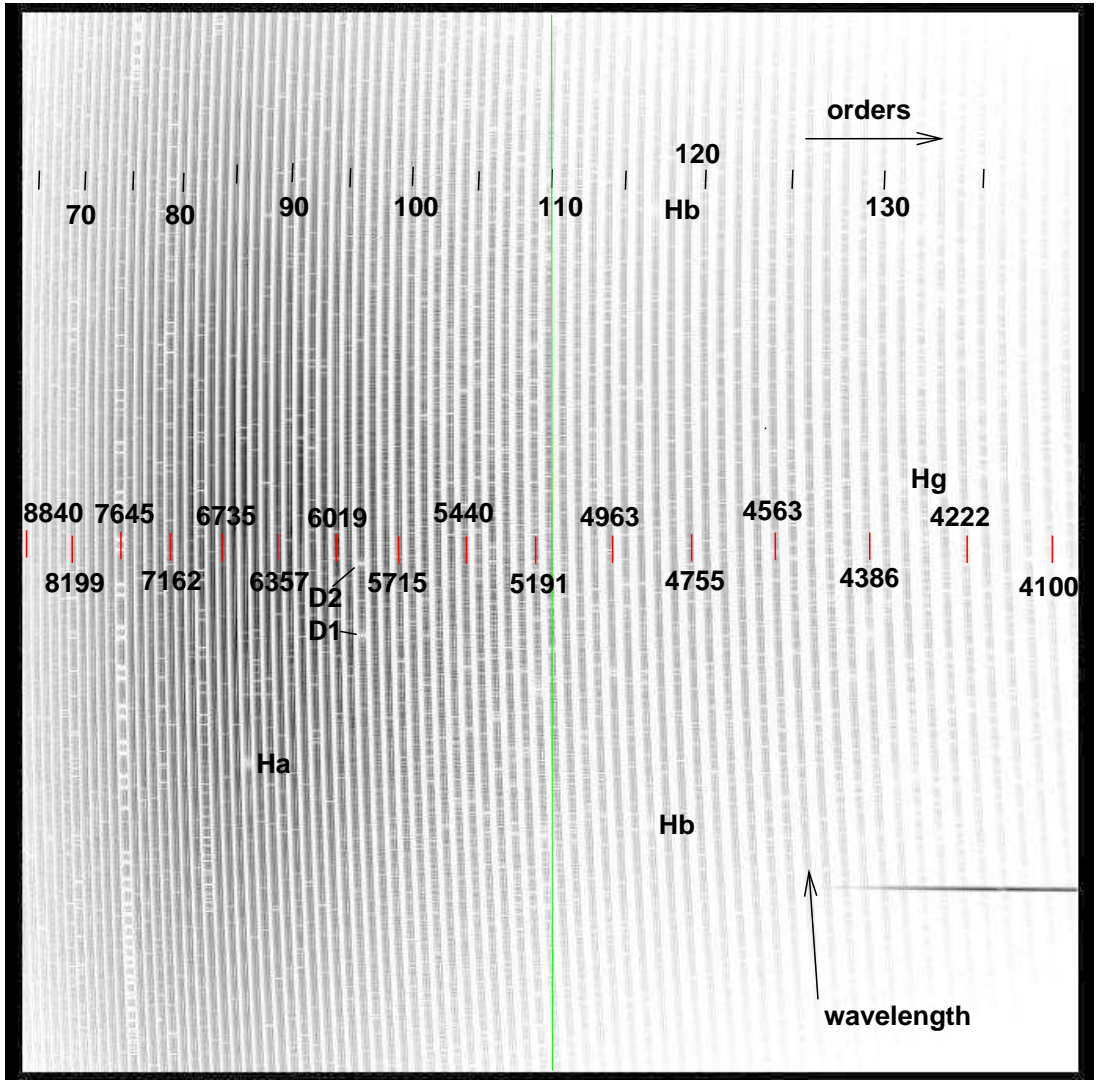}
\caption{\label{fig:orders} Order map of CHIRON. The order numbers are
  indicated on the top, central wavelengths of some orders and several
  characteristic  lines  are overlayed  on  the low-contrast  negative
  display of the $\alpha$~Cen~A spectrum.}
\end{figure*}

To  set the  top  level  requirement for  resolution,  we carried  out
simulations of  photon-limited Doppler precision and found  that a SNR
of  200 and  resolution of  $R  =80\,000$ yielded  internal errors  of
1.2\,\ms.   Increasing the simulated  resolution to  120\,000 improved
the single measurement  precision in our simulations for  SNR of 200
to  1.0\,\ms~    \citep[see  also][]{Bouchy01}.   However, the  increased
spectral  resolution  then  required  exposure times  longer  than  15
minutes in  order to reach  a SNR  of 50 for  the typical star  on our
program.  Thus,  we optimized the instrumental  resolution for Doppler
precision, with the desired SNR as limiting consideration.

A high premium  was placed on the instrument  stability, especially on
stability in  the SLSF \citep{Spronck13b,  Spronck12a, Spronck12b}, so
some  thermal  control was  included  and  light  was coupled  to  the
instrument  with  a  fiber.   CHIRON  is  one  of  the  few  fiber-fed
spectrometers using an  iodine cell.  Finally, we wanted  to reach the
highest possible  efficiency to  offset the small  telescope diameter.
With a fixed spectral format, CHIRON covers a broad spectral range and
is useful for a wide variety  of science programs.  We made use of the
existing acquisition/calibration/guiding  module deployed in  2008 for
the fiber echelle.

The  instrument  concept  was  developed  in  2009.   Its  design  and
construction were completed in  one year and science observations with
CHIRON began  in March 2011. During  the first year  of operation, the
internal RV measurement precision  was about 0.8~\ms, but the velocity
rms  scatter for  chromospherically stable  stars was  generally about
2~\ms.  To further improve the RV precision, throughput and stability,
the instrument  was upgraded between  January and May 2012  in sereval
respects.   The old  echelle  grating  was replaced  by  a new  higher
efficiency R2  Richardson grating mounted  in a vacuum  enclosure.  We
developed  a  sol-gel coating  facility  to  apply an  anti-reflective
coating to some optics (the prism, echelle vacuum enclosure window and
other  optics). We  improved thermal  control; an  exposure  meter for
photon-weighted exposure timing was  installed. The CCD controller was
replaced,  and the  input  fiber was  changed  from a  circular to  an
octagonal fiber for better modal scrambling of the light.

The instrument  is described in  \S\ref{sec:inst}. In \S\ref{sec:data}
the observing procedure and  data reduction are briefly covered, while
actual  instrument performance is  discussed in  \S\ref{sec:perf}. The
paper  closes   with  conclusions  in   \S\ref{sec:concl}.  Additional
technical  information on  the  instrument  can be  found  on its  web
site.\footnote{http://www.ctio.noao.edu/noao/content/chiron}

\section{Instrument description}
\label{sec:inst}

CHIRON  is located  in the  Coud\'e room  of the  1.5-m  telescope. It
receives  the light  from a  star (or  a calibration  lamp)  through a
multi-mode  fiber, which  is permanently  installed at  the telescope.
The  optical configuration and  spectral format  of CHIRON  are fixed,
covering  the wavelength  range from  415\,nm to  880\,nm in  a single
exposure. The  fiber image  can be transformed  by a slicer  or masks,
offering some  flexibility.  In the  nominal mode with  slicer, CHIRON
reaches spectral resolution of $R=79\,000$ with 3-pixel sampling. Each
CCD  pixel  corresponds to  1.075\,km~s$^{-1}$  in  RV.  For  reader's
convenience main  characteristics of the instrument  are summarized in
Table~\ref{tab:par};  they are  discussed further  in the  paper.  


\begin{table}[ht]
\center
\caption{Main parameters of CHIRON}
\label{tab:par}
\medskip
\begin{tabular}{  l    }
\hline
$R = \lambda/\Delta \lambda$: 27\,400, 79\,000, 95\,000, 136\,000 \\
Wavelength coverage:  415\,nm to 880\,nm \\
Spectral orders:  138 to 66 \\
Collimator:  $F=600$\,mm, beam diameter 130\,mm \\
Grating:  $63.9^\circ$ blaze, 31.6\,l/mm, 130$\times$260\,mm \\
Cross-disperser:  LF7 prism, apex $62^\circ$, one pass \\
Camera: oil triplet $F=1012\,$mm, $D=140$\,mm \\
CCD: 4096(H)$\times$4112(V), 15\,$\mu$m pixels, graded-AR \\
Fiber feed: 100\,$\mu$m octagonal core, $2.7''$ on the sky \\
\hline
\end{tabular}
\end{table}

\subsection{Optical design}

\begin{figure}[ht]
\epsscale{1.0}
\plotone{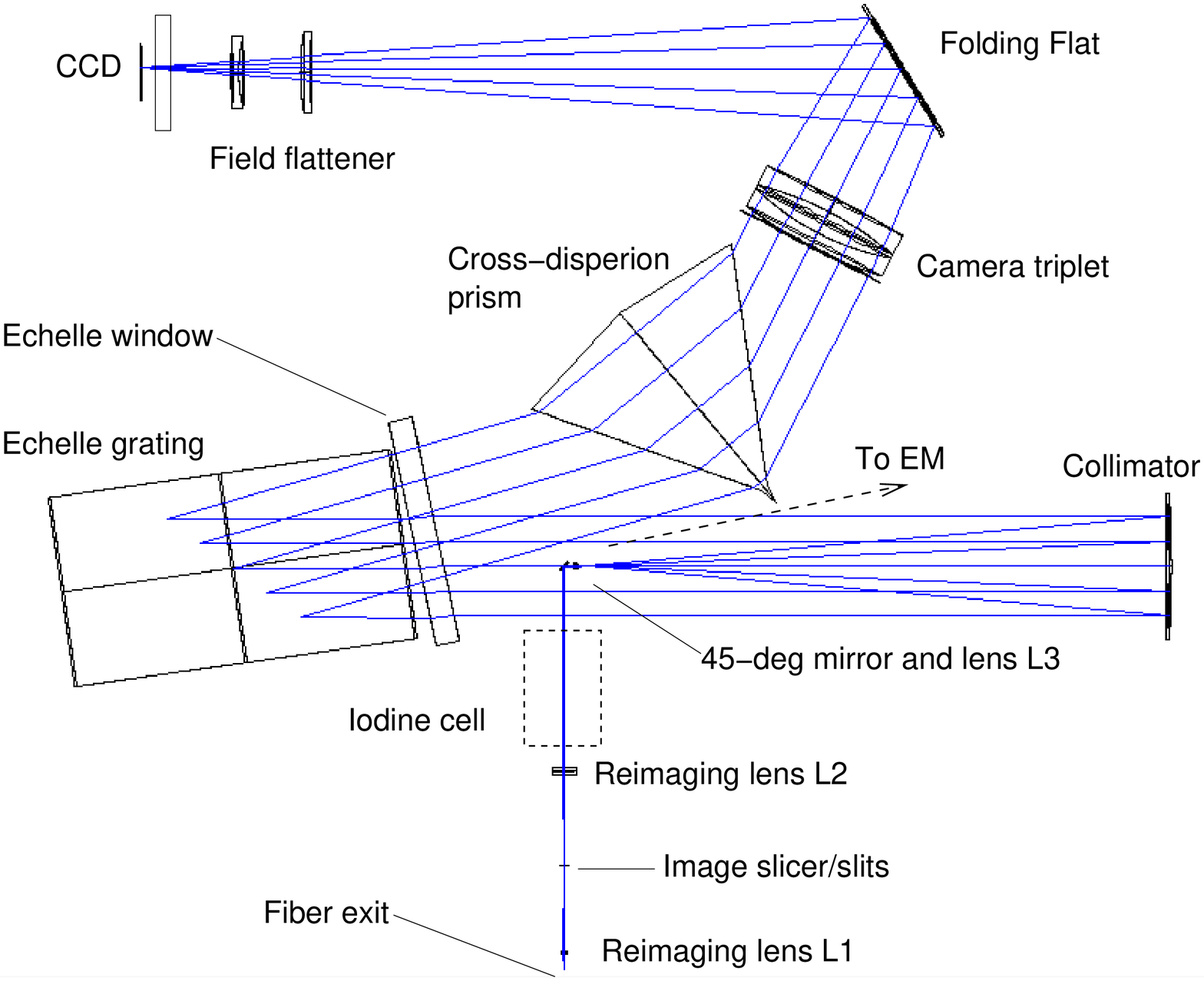}
\caption{\label{fig:opt}
Layout of the CHIRON optics (see text).
}
\end{figure}

Most  modern  high-resolution  echelle  spectrometers  belong  to  two
families.  One of these is the double-pass Littrow design, as e.g.  in
SOPHIE \citep{SOPHIE} or PFS  \citep{PFS}.  Another is the white-pupil
design  pioneered   by  ELODIE   \citep{ELODIE}  and  used   in  FEROS
\citep{FEROS}, HARPS  \citep{HARPS}, HERMES \citep{HERMES},  and other
instruments. CHIRON has a  pseudo-Littrow design; the beams before and
after the echelle grating are separated by a relatively large angle of
$2\gamma =  11^\circ$ and go  through different optics,  resembling in
this respect  HERCULES \citep{HERCULES}, the Blanco  Echelle, and some
other classical spectrographs. This choice is driven by the simplicity
and the  desire to minimize  the number of optical  elements, boosting
the  efficiency; the  loss  of diffraction  efficiency  is only  $\cos
\gamma = 0.995$ \citep{Schroeder}.   A Littrow design with double pass
through  refractive optics and  prism would  have twice  the chromatic
aberration and a variable slit tilt along the orders.

The drawbacks  of using an  incidence angle $\gamma=5.5^\circ$  in the
cross-dispersion direction  are the  elliptical beam footprint  on the
camera  and the  rotation of  the slit  image by  $24^\circ$.  This is
compensated by appropriate counter-rotation of the entrance slit, at a
cost of $\cos 24^\circ = 0.91$ loss in resolution.

The light beam emerging from  the fiber is transformed by small lenses
and  collimated by the  150-mm $F/4$  on-axis parabolic  mirror before
reaching  the echelle grating  (Fig.~\ref{fig:opt}). The  nominal beam
diameter on the collimator is 130\,mm. We used an old large R2 echelle
during the first year after commissioning.  However this grating had a
measured  peak efficiency of  about 55\%.   This grating  was replaced
during the CHIRON upgrade with a more efficient grating with protected
silver
coating.\footnote{http://gratings.newport.com/information/techdata/Coating\_Reflectivity.pdf}
The  new  grating  was  procured from  Richardson  Labs/Newport:  31.6
lines/mm,  nominal  blaze  angle  63.9$^\circ$, Zerodur  substrate  of
135$\times$265$\times$45mm.   The  ruled  area is  130$\times$260\,mm.
The measured efficiency at blaze  peak at $\lambda = 514$\,nm is 82\%.
The 130-mm beam is slightly vignetted by the grating.

The  new  echelle   grating  is  housed  in  a   vacuum  enclosure  to
significantly reduce variability in the temperature or pressure of the
air layer in  contact with the grating that  would otherwise introduce
changes in  the dispersion. Light  passes twice through  the enclosure
window,  made  of  BK7  glass   with  a  $1.5^\circ$  wedge  to  avoid
reflections. Nominal surface quality of the window is $\lambda/4$ with
$<$4  fringes  power.   Reflection   losses  at  window  surfaces  are
minimized by  the sol-gel anti-reflection  (AR) coating on  the window
($R<0.25$\% per  surface at 500\,nm). The reflectivity  of the sol-gel
coating  increases to  2\% at  the  blue and  red ends  of the  CHIRON
wavelength range.  Double reflection from the window produces a bright
feature in the  blue part of the spectrum which  is mostly outside the
free  spectral range.  The vacuum  echelle enclosure  has  20-mm thick
walls  to minimize  its  mechanical reaction  to changing  atmospheric
pressure.

The cross-dispersion prism was made out of a Schott LF7 glass blank by
the TORC  company in Tucson. This  glass was selected  as a compromise
between cost, availability, and uniformity of inter-order spacing. The
prism apex  angle is $62^\circ$, side length  260\,mm, height 160\,mm.
Only  one pass  through  the  prism located  {\em  after} the  echelle
grating  ensures a constant  slit tilt  along the  orders; this  is an
advantage  over  the Littrow  design  with  a  prism in  double  pass.
Inhomogeneity of the refractive  index produces some aberration of the
transmitted  beam,  despite  an  attempt  to correct  them  by  manual
retouching of  the prism.  During  the upgrade, the prism  was sol-gel
coated to optimize transmission at 500\,nm and at $54^\circ$ incidence
angle (reflection loss at  each surface $R<0.6$\% at 550\,nm, $<2.5$\%
over  entire  wavelength  range).   The  internal  glass  transmission
averaged over the beam is 98\%.

\begin{figure}[ht]
\epsscale{1.0}
\plotone{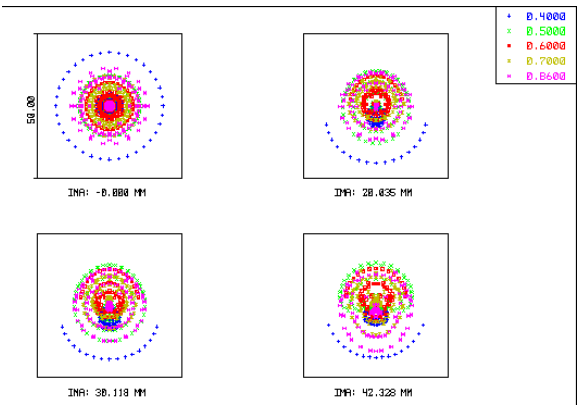}
\caption{\label{fig:APO}  Spot  diagrams  of  the APO-140  triplet  in
  single pass. The size of each square is 50\,$\mu$m. }
\end{figure}

The camera is a commercial  triplet lens APO-140 produced by Telescope
Engineering Company\footnote{http://www.telescopengineering.com}. This
oil-spaced triplet  lens has a  light diameter of 140\,mm.   The focal
length with a 2-element  field flattener is 1012\,mm. The manufacturer
kindly provided the optical  prescription, allowing us to evaluate the
lens  performance. The spot  diagrams fit  within 50\,$\mu$m  over the
whole 60$\times$60\,mm  square field (Fig.~\ref{fig:APO}).   The major
residual   aberration   is  chromatic   defocus,   so   the  lens   is
diffraction-limited  at   any  given  wavelength.    The  triplet  and
flattener lenses have  broad-band AR coatings, with a  typical loss of
0.25\% per surface.   The beam after the camera is  folded by a 150-mm
flat mirror custom-coated to  97\% reflectivity, making the instrument
more compact.

The wavelength range of CHIRON avoids the near-UV region (e.g. calcium
H and  K lines) because  it would require a  different custom-designed
camera  lens  and  because  the  fiber transmission  degrades  in  the
blue. As the iodine lines  are concentrated in the 500--600\,nm region
of the spectrum,  the loss of UV does not compromise  the main goal of
the instrument.

\subsection{Image slicer}

Achieving high spectral resolution with  a single fiber is possible by
decreasing the  fiber diameter (with associated light  loss at input),
increasing  the beam  size (and  increasing the  instrument  cost), or
using an R4 echelle (more orders  on the detector).  We have chosen to
slice the  fiber image  in 3  segments, gaining a  factor of  three in
resolution over the bare fiber without much light loss.  CHIRON uses a
Bowen-Walraven  image slicer.   For fabrication  reasons,  two mirrors
replace the  standard design with  total internal reflection,  as used
e.g.  in FEROS \citep{FEROS}. Mirrors with high reflectivity and sharp
edge are available at  low cost. As the three slices have  0, 2, and 4
reflections respectively, the light loss is acceptably small (measured
efficiency  0.82).   The  slicer   is  described  in  more  detail  by
\citet{CHI-2010,CHI-2012}.

To reduce  the defocus inherent to  this image slicer,  the $F/5$ beam
emerging from  the fiber is  transformed to a  slow $F/41$ ratio  by a
small lens  L1 (Fig.~\ref{fig:opt}). The fiber image  is magnified 8.3
times,   to  0.825\,mm   diameter;  the   sliced  ``slit''   image  is
0.27$\times$2.4\,mm.  After the slicer,  a second  lens L2  of 100\,mm
focal length collimates the beam and, at the same time, forms the pupil
image close to the following  elements: the tiny diagonal mirror and a
$F=12.5$\,mm  lens L3 that  couples the  beam to  the  collimator.  These
elements are  located near the  collimator focus and supported  by two
thin vanes.

The  reflective slicer  is tunable  and offers  extra  flexibility. By
moving it  out of the beam,  the un-sliced fiber image  is passed into
the  instrument.  The  spectral  resolution is  thus  degraded to  $R=
27\,400$, but the fiber projects  to a smaller number of binned pixels
on the detector, increasing  the signal-to-noise ratio (SNR) for faint
stars.   The slicer  unit also  holds two  slit masks  of  0.2\,mm and
0.1\,mm width that  can block the fiber image  by the same translation
motion of the slicer unit.  The wider slit provides similar resolution
as  the  slicer, but  with  narrower  and  ``cleaner'' orders  in  the
cross-dispersion direction.  The narrow  slit brings the resolution to
$R=136\,000$ at the expense of  even larger light loss. These two slit
masks  are intended for  observing very  bright stars  with an  aim of
achieving the highest RV precision. The four slit modes are summarized
in  Table~\ref{tab:modes}.   The  resolution  is determined  from  the
measured  FWHM of  the line-spread  function, SLSF.  The  last columns
gives efficiency of each mode relative to the bare fiber.

\begin{table}[ht]
\center
\caption{Basic observing modes of CHIRON}
\label{tab:modes}
\bigskip
\begin{tabular}{  l  l l l }
\hline
 Mode & binning    & Spectral & Relative \\
      & H$\times$V & resolution       & efficiency \\
\hline
 Slicer & 3$\times$1  & 79\,000 & 0.82 \\
 Slit  & 3$\times$1 & 95\,000 & 0.25 \\
 Narrow    & 3$\times$1 & 136\,000 & 0.11  \\
 Fiber  & 4$\times$4 & 27\,400 & 1.0\\
\hline
\end{tabular}
\end{table}

\subsection{Iodine cell}

For precise RV work, the iodine  cell is inserted in the narrow (3\,mm
diameter) collimated  beam after the  second lens L2,  without causing
defocus. The cell  is a glass cylinder of  50\,mm diameter and 100\,mm
length filled  with iodine  vapor.  Its windows  are AR-coated  on the
external surfaces.  It  is placed in a heated  and insulated container
and  maintained at  $+40^\circ$C  stabilized temperature  to keep  all
iodine vaporized.   Owing to the  insulation, the 10-W  heater working
typically  at half-power  is sufficient  to maintain  a  constant cell
temperature.

It would  be preferable to place  the I2 cell {\em  before} the fiber,
but  the restricted space  made this  option unfeasible.   In CHIRON,
optical  aberrations of  the cell  may influence  the  SLSF (sometimes
called point spread  function or PSF) or shift  the spectrum, compared
to  the non-iodine configuration.   This is  relevant for  the stellar
template spectra used in forward modeling of the Doppler analysis (see
\S\ref{sec:iodine}).   Fortunately, the small  beam diameter  helps in
reducing the aberrations of the cell to an undetectably small level.

After  double reflection from  the front  and back  sides of  the cell
windows,  the light  interferes with  the main  (transmitted) parallel
beam   and  modulates   the  transmitted   spectrum   with  sinusoidal
``fringes'' that have a  relative amplitude of $\sqrt{r_1 r_2}$, where
$r_1$ and  $r_2$ are intensity reflectivity coefficients  of the glass
surfaces.   The fringe  period is  $\lambda^2/(2 d  n)$  at wavelength
$\lambda$ for a window of thickness $d$ and refractive index $n$.  Two
fringe systems are produced by the two cell windows; very fine fringes
resulting  from the interference  between the  windows can  usually be
neglected.

Fringing  is intrinsic  to  {\em all}  iodine  cells.  However,  under
certain conditions  (diverging or  inclined beam, wedged  windows) the
modulation is averaged out and  becomes acceptably small.  This is the
case for  CHIRON where  the cell  has 6-mm windows  and does  not show
evidence of  fringing. We observed fringing (modulation  of the quartz
spectra) when other cells with 3-mm windows were placed in CHIRON.

\subsection{Mechanical and thermal aspects}

\begin{figure}[ht]
\epsscale{1.0}
\plotone{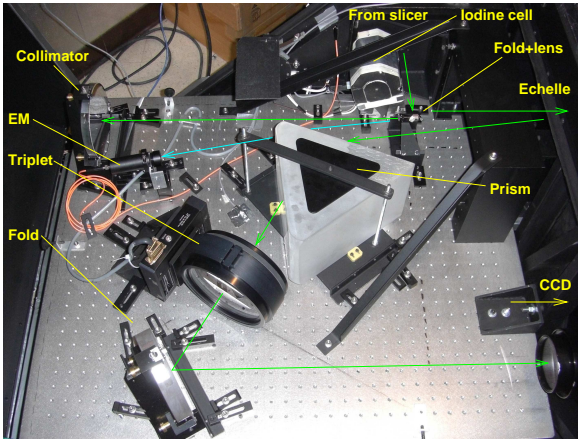}
\caption{\label{fig:inside} Optical elements of CHIRON and the light path.}
\end{figure}

All optical elements of the spectrograph are mounted on a standard optical table 
from Thorlabs with a size of 90$\times$75$\times$11\,cm (Fig.~\ref{fig:inside}). 
The vacuum echelle enclosure and the fore-optics are mechanically attached to 
the table but extend beyond its perimeter.

The table  is oriented horizontally  and connected at three  points to
the steel frame;  deformations of the frame should  not be transmitted
to the  optics.  The  exception is  made for the  CCD dewar,  which is
attached to the frame.  The dewar  is a source of instability owing to
its  variable mass and  colder temperature,  therefore it  is isolated
from  the main  optics  mechanically, electrically,  and thermally.  A
light-  and air-tight  enclosure  is integrated  with  the frame.  The
enclosure lid  can be  opened for  easy access to  the optics.  On the
outside,  the enclosure is  covered by  thermal insulation  (panels of
polyisocyanurate foam, thickness 25\,mm).

Stabilization   of  the  instrument   temperature  is   important  for
minimizing  displacement of  the  spectrum on  the  detector caused  by
temperature-related  mechanical   deformations  and  by   the  thermal
expansion of  the grating. As  part of the  upgrade, CHIRON now  has a
second stage  of insulation.  The insulated  spectrometer enclosure is
surrounded by a ``warm room'' constructed from an aluminum 80/20 truss
and  covered  with  layers  of  PRODEX,  a  lightweight  and  flexible
polyethylene foam covered by aluminum  foil on both sides. The thermal
room  has forced  air circulation  and air  temperature  stabilized at
$+22^\circ$C  by a controlled  heater. The  temperature was  chosen to
match the maximum environmental  temperature in the Coud\'e room. This
creates a  year-round stable  environment de-coupled from  the outside
temperature variations.  The  inner high-precision temperature control
of  the instrument  is  done with  a  12.5\,W internal  heater in  its
support  structure,  a  sensor,  and the  Lakeshore  controller.   The
set-point  is  $+23^\circ$C.  Owing  to  good  thermal insulation  and
stable outside air temperature, only a few watts of heating are needed
to  stabilize CHIRON's  internal temperature;  low and  stable heating
power also means small and stable internal temperature gradients. Most
of the heat dissipates through the enclosure walls and by conductivity
to  the dewar,  which is  not perfectly  isolated from  the structure.
Temperature  is monitored at  four points  inside the  instrument.  We
found that internal temperature variations are five times smaller than
the  variations of  the surrounding  air  in the  thermal room.   When
thermal control  operates normally, the  rms fluctuations at  all test
points are about 0.1$^\circ$C.

As CHIRON  is stationary, the gravity-induced  deformation is constant
and irrelevant.  However, the  dispersion direction coincides with the
mechanically weak vertical axis  of the table.  A vertical temperature
gradient would  bend the table and  move the spectrum.   We measured a
systematic shift  of the spectrum  with temperature of  1.9\,pixel per
degree in the dispersion direction  and $-0.7$ pixel per degree in the
cross-dispersion direction. In  normal operation the spectrum position
on  the detector  is  stable to  a  small fraction  of  pixel in  both
directions.  Dewar refill shifts the  spectrum by 0.07 pixels and that
motion is easily tracked by the iodine reference.

A  pressure   sensor  is  installed  to  monitor   variations  of  the
atmospheric pressure.   This is not  as critical since the  echelle is
now in a  vacuum enclosure. The pressure inside  the echelle enclosure
is  also monitored.  It  grows slowly  with  time, mostly  due to  the
out-gassing.   The enclosure  is  pumped once  the  pressure inside  it
approaches 1\%  of atmospheric  pressure, 8\,mB, currently  once every
five or six months.

CHIRON has two remotely controlled motors, one for focusing (moves the
APO-140 triplet mounted on a linear translation stage) and another for
moving  the slicer  unit  to change  the  slit mode.  Both motors  are
manufactured  by  Physik Instrumente  and  are  driven  by the  EZSV23
single-board  units\footnote{www.allmotion.com}  that  connect to  the
computer  by serial  lines.  The  translation of  the  iodine cell  is
performed by a simple linear actuator with 50-mm range.

\subsection{CCD detector} 

The  CHIRON  detector  is  a   CCD  device  CCD231-84  from  e2v  with
4096$\times$4112 square  15\,$\mu$m pixels. It has  a gradient coating
in the  line direction  for optimum sensitivity  over a  wide spectral
range. For  this reason we oriented  CCD columns parallel  to the main
(echelle) dispersion,  with cross-dispersion along the  lines. The CCD
is  housed in  a  dewar with  liquid  nitrogen cooling  and a  nominal
holding time  of 36 hours. It  is refilled daily  during daytime.  The
CCD temperature is stabilized using a Lakeshore controller. 

After CHIRON commissioning  in 2011, the detector was  operated with a
provisional controller which  was replaced in 2012 by  the new Torrent
controller developed at NOAO \citep{Torrent}. To our knowledge, CHIRON
is the first astronomical  instrument using Torrent. The controller is
connected to the dewar by two cables of 40\,cm length.

The   Torrent  controller  works   with  a   pixel  readout   rate  of
129\,kHz. Readout of the full  un-binned chip with all four amplifiers
takes $\sim$35\,s.  We normally use the CCD with 3$\times$1 (binned in
the cross-dispersion direction)  or 4$\times$4 binning, which shortens
the readout  time to 18\,s and  5\,s, respectively. The  gain is about
1.3 electrons per  ADU.  The readout noise of 5.4  to 5.7 electrons in
this mode is higher than the  intrinsic CCD noise, being affected by a
periodic component  which shows  as a ``fringe''  pattern in  the bias
frames and originates in the controller itself.

Linear response of  the CCD is important for  precise spectroscopy. In
this respect, our  CCD and controller combination is  excellent in the
full  signal range  up to  65\,kADU.  The  ratio of  exposure  time to
counts from a stabilized diffuse  light source is constant to 0.5\% or
better.  We developed an  alternative method  to investigate  gain and
linearity   from  the   ratio  of   two  images   (e.g.   quartz  lamp
spectra).  This method  does not  require  flux stability  and can  be
applied  to calibration  spectra  at any  time.   The charge  transfer
efficiency is  yet another critical  parameter for precise RV.   It is
excellent in both line and  column directions; we can only place upper
limits on the charge spread of $<10^{-5}$ per transfer.

\subsection{Fiber feed and guider}

\begin{figure}[ht]
\epsscale{1.0}
\plotone{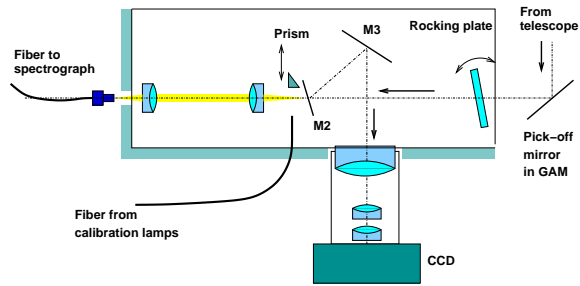}
\caption{Fiber feed of CHIRON.
\label{fig:feed}
}
\end{figure}

The stellar light collected by  the telescope is reflected towards the
CHIRON fiber by a diagonal pick-off mirror. This mirror is attached to
the  guide probe  inside the  regular guiding  and  acquisition module
(GAM) of the 1.5-m telescope.  By placing the probe on-axis, we direct
the  light  to  CHIRON.    The  fiber  feed  (Fig.~\ref{fig:feed})  is
permanently  available  along with  the  standard on-axis  instruments
installed at the telescope.  The observing time can be shared flexibly
between CHIRON and these instruments.

The CHIRON fiber feed module is extremely compact, being restricted by
the space available on the side  of the GAM. It can be easily detached
from the telescope without disconnecting fibers or cables. This is the
only way to access the module for alignment and service.

Originally, we used  a 15-m fiber with a  100\,$\mu$m round core (type
FPB1001120140  from Polymicro).  However,  better light  scrambling is
achieved     by     a     fiber     with     an     octagonal     core
\citep{Chazelas10,Spronck12a}  and  this was  installed  in May  2012.
This  fiber was  made by  CeramOptec and  has a  100\,$\mu$m octagonal
core, 660\,$\mu$m  round cladding,  acrylate jacket, length  of 20\,m,
numerical aperture 0.22$\pm$0.02, and FC connectors on both ends.  The
fiber has low focal ratio degradation (FRD); with F/5 input and output
at  514\,nm the transmission  is about  0.87. We  found that  only $<2
\times  10^{-5}$ of  the incoming  light is  scattered into  the fiber
cladding  (this   fiber  replaced  another  octagonal   fiber  with  a
silicon-type jacket that resulted in excessive light loss).

In  CHIRON,  we image  the  star  onto the  fiber  input  end.  It  is
preferable to imaging the  telescope pupil because fibers scramble the
light spatially  (near-field) better  than in angle (far  field). The
far-field  (pupil   image)  is  more  stable,   while  the  near-field
illumination is  affected by guiding  errors and hence needs   better
scrambling \citep[see  e.g.][]{Bouchy12}.  The star  is re-imaged onto
the fiber by two small lenses which transform the beam from $F/7.5$ to
$F/5$  (Fig.~\ref{fig:feed}). The 100\,$\mu$m  fiber core  projects to
$2.7''$ on the  sky.  The fiber is conjugated to  the hole of 0.15\,mm
diameter  in  the concave  and  tilted  mirror  M2 that  reflects  the
surrounding field  to the acquisition  and guiding camera.  Behind the
concave mirror, a 2-mm prism can  be moved into the beam to direct the
light  of fiber-coupled  comparison lamp  into the  spectrograph.  The
prism is driven by a miniature motor HS-85 from Hitek.

The guiding  camera contains an  un-cooled CCD GC\,650  from Prosilica
with  693$\times$493 pixels,  which  project at  $0.42''$ scale  (3$'$
field)  on the sky.   The guiding  is normally  done by  measuring the
centroid of the halo of stellar  light around the hole. In the case of
binary stars such as $\alpha$~Cen, we can guide on one component while
the other is placed in the hole.

Originally, the  correction signals were sent to  the telescope drives
every second.   For a fixed exposure  time, the average  signal on the
CCD from the same star varied by a factor of two or more.  We strongly
suspected that poor guiding was  the main reason for these variations,
as the fiber size of $2.7''$  is large enough to admit most light even
under poor  seeing.  Backlash of  the telescope drives,  especially in
declination, could be the  most serious problem affecting the guiding.
In April 2013, we installed  an improved guiding system to correct the
pointing more  accurately with a tilted  transparent ``rocking plate''
in front of the focus.

The rocking plate  is a 6\,mm thick AR-coated  window, with a diameter
of 40\,mm.   Its tilt in two  orthogonal directions is  actuated by two
miniature     stepper    motors\footnote{AM1020-2R-V-12-250-23    from
  http://www.faulhaber.com}.  The full range  of the plate tilt, $\pm
15^\circ$,  provides  an image  displacement  corresponding  to $  \pm
9\farcs5$ on the sky. The plate position is currently updated at about
2\,Hz. When the plate angles reach a programmable limit, the telescope
is commanded to move so that  the rocking plate will move back towards
the  middle  of its  range.   The  average  gain in  the  spectrograph
throughput  was measured  over several  nights  to be  23\% after  the
addition of the tip/tilt module and cleaning of the pick-off mirror.

Calibration lamps,  thorium-argon (ThAr) and quartz, are  mounted in a
separate  module and  coupled  by a  fiber  with 0.4\,mm  core to  the
front-end module.   The cathode of the  ThAr lamp is  imaged onto that
fiber by a small lens.  A  tilted un-coated glass wedge in front of the
lens  couples the  fiber to  the quartz  lamp (without  moving parts).
Spatially separated reflections from the wedge allow us to balance the
spectrum  of the lamp  by selective  attenuation of  its blue  and red
parts.   One side  of  the  wedge projects  the  quartz lamp  directly
through the 4-mm  thick green filter BG38.  Reflection  from the other
side of the wedge feeds attenuated unfiltered light from the same lamp
(mostly red), allowing calibration  of orders over the full wavelength
range in a single quartz exposure.

\subsection{Exposure meter}

\begin{figure}[ht]
\epsscale{1.0}
\plotone{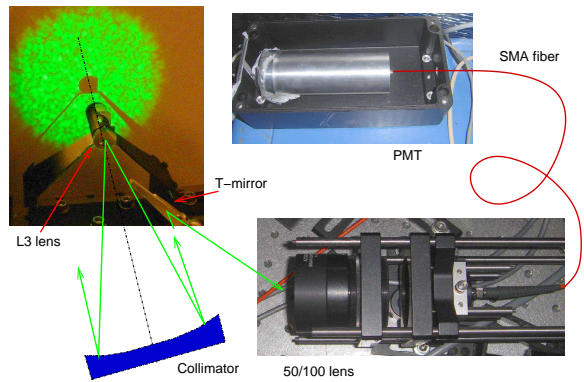}
\caption{Exposure meter.
\label{fig:EM}
}
\end{figure}

An exposure meter  was installed as part of  the instrument upgrade to
calculate   a   photon-weighted  midpoint   for   the  exposure   time
\citep{Kibrick06}. This is critical for proper calculation and removal
of  the  barycentric velocity  during  Doppler  planet searches.   The
exposure meter system at  Lick Observatory was originally developed to
optimize   exposure   times   for   observations   with   photographic
plates. This system uses a  reflective propeller blade to divert about
8\%  of the  light from  behind the  slit to  a  photo-multiplier tube
\citep{Kibrick06}.   The Keck  exposure  meter has  a similar  design,
including a  propeller blade  to pick off  light behind the  slit.  At
CHIRON,  we elected  to pick  off about  1\% of  the  collimated light
before it reaches the grating and direct it to the exposure meter (EM)
by an  optical fiber. Originally, the  15-mm central part  of the beam
(which is lost anyway to the central obstruction) was reflected to the
EM by  a circular  mirror with  a hole.  However,  the beam  center is
``dark''  because of  the central  obstruction in  the  telescope (the
fiber  preserves angles,  so that  the secondary
mirror  obstruction is  imprinted in  the outgoing  beam).  Therefore,
during  the  2012 upgrade,  the  circular  mirror  was replaced  by  a
triangular mirror in front of  the vane that supports the small optics
near the collimator focus (Fig.~\ref{fig:EM}).  The base of the mirror
is 5-mm  wide, its length  is 50\,mm. With this  repositioned pick-off
mirror and  larger area,  the EM samples  about 2\% of  the collimated
beam.

The portion of the parallel beam reflected by the EM mirror is focused
by a lens  ($D=50$\,mm, $F=100$\,mm) on the SMA  fiber with 200-$\mu$m
core. The fiber delivers  light to the photo-multiplier model H9319-11
from Hamamatsu. A filter of  545/90\,nm restricts the measured flux to
the  ``iodine''  region  of  the  spectrum.   The  photon  counts  are
accumulated  with  0.1\,s  time  cadence  and transmitted  to  the  EM
computer through  a serial interface.  A $V=5^m$ star gives  a typical
count rate of $10^3$\,s$^{-1}$ in the EM.

The  primary function  of the  EM  is to  calculate the  flux-weighted
midpoint time of exposure, as  needed in the precise RV work. However,
the  EM can also  be used  to auto-terminate  exposures after  a given
amount of accumulated EM counts  are acquired, instead of setting fixed
exposure times. This contributes to a more uniform data quality.

\subsection{Computers and software}

CHIRON is controlled by three PC computers running under Linux. One is
used  for   taking  images,  for  the  instrument   control,  and  for
environment monitoring.  The CHIRON  control software was developed by
M.~Bonati. All instrument  functions (slicer, focus, comparison lamps,
iodine cell  movements, etc)  are automated.  The  operator can  use a
graphical user  interface (GUI) for setting  instrument parameters and
performing  exposures.    Command-line  mode  and   scripts  are  also
available.  The  scripting   enables  automatic  calibrations  at  the
beginning  and end  of the  night, setting  observing modes  by a single
command,    and   interaction    with    the   high-level    software
\citep{Brewer13}. 

The second computer is responsible for the guiding. It is connected to
the GC-650 CCD  camera by a reserved Gigabit Ethernet  line.  The guider
software  controls  the  stepper  motors  and sends  commands  to  the
telescope. The third computer serves the exposure meter. 



\section{Observations and data reduction}
\label{sec:data}

\subsection{Observing procedure}

The four  slit modes of CHIRON  and associated CCD  binning are listed
above in Table~\ref{tab:modes}. We do  not change the readout speed of
the  CCD.  Each  mode is  set by  a script.  More complex  scripts are
available to automatically  execute sequences of calibration exposures
before and after the night. In addition to the slit mode, the operator
also must specify  the object name, program identifier,  and the IN or
OUT position of the iodine  cell.  All these parameters are entered in
the GUI.

The telescope  is slewed  to an  object, the star  is centered  on the
fiber using  the guiding/acquisition camera,  and the guiding  loop is
closed. At this point, one or several exposures are taken, accompanied
if needed by calibration spectra.

Recently, a high-level software for observation planning and execution
was implemented to further  automate the process \citep{Brewer13}. The
observing program for  the night is prepared by  the scheduling center
at  Yale and  sent to  the local  computer at  CTIO, where  it  can be
accessed by a browser. The  operator selects an object from the target
list to observe and the  software then sends the target coordinates to
the telescope control system, sets up the requested observing mode, and
populates all fields in the data-taking GUI. This eliminates potential
human  errors,  leaving  to   the  operator  the  important  tasks  of
evaluating  observing  conditions,   monitoring  the  instrument,  and
reacting to non-standard situations.

After each night, the data are transferred to Yale for quality control
and processing.   A web-based system  for internal data  quality check
has been developed. In parallel, several parameters are extracted from
the CHIRON log files and FITS  headers and used to update the web page
that  displays instrument temperatures,  pressure, fluxes,  and tracks
the resolution  and stability from  the ThAr spectrum position  on the
CCD.

\subsection{Extraction of spectra}

\begin{figure}[ht]
\epsscale{0.7}
\plotone{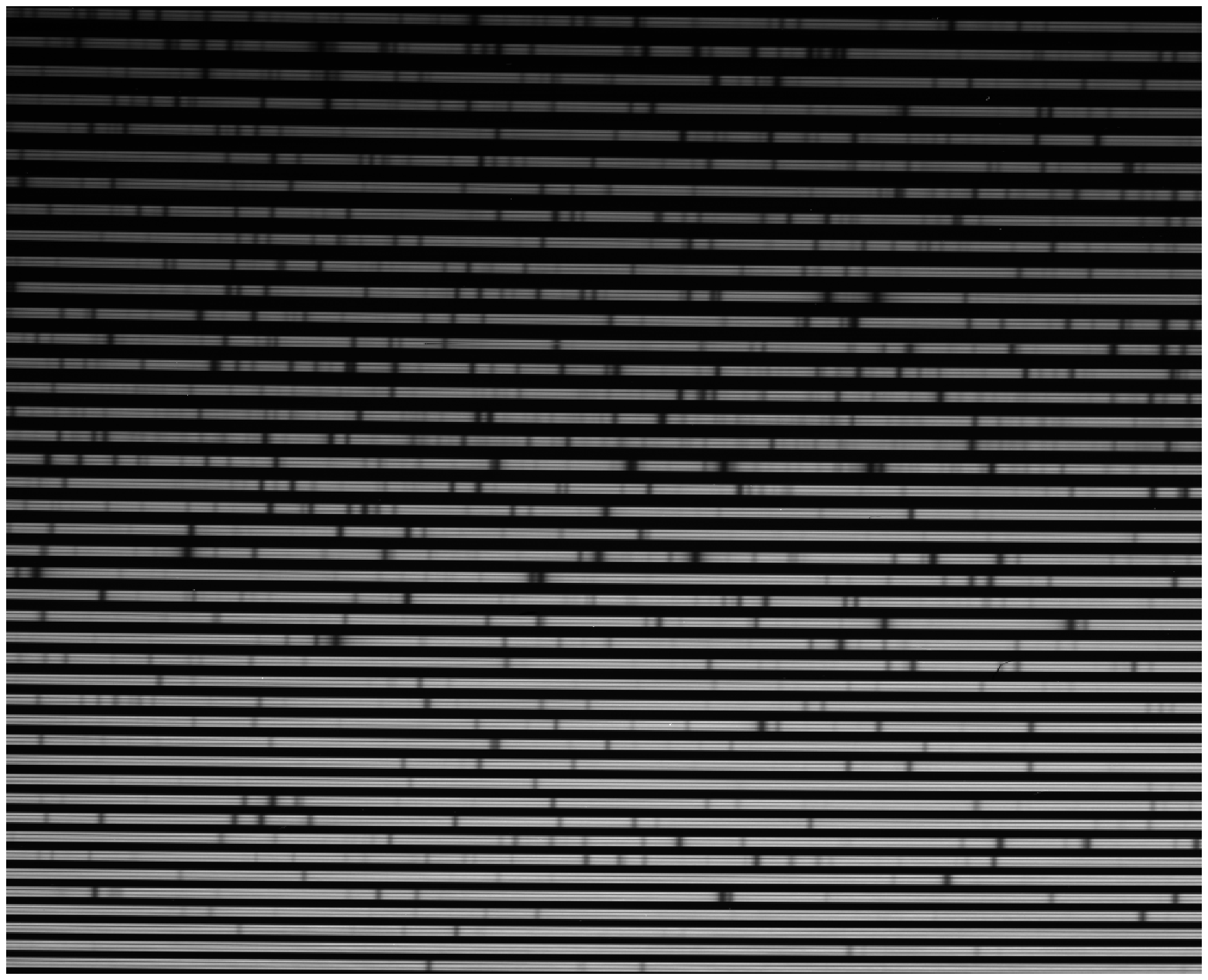}
\epsscale{1.0}
\plotone{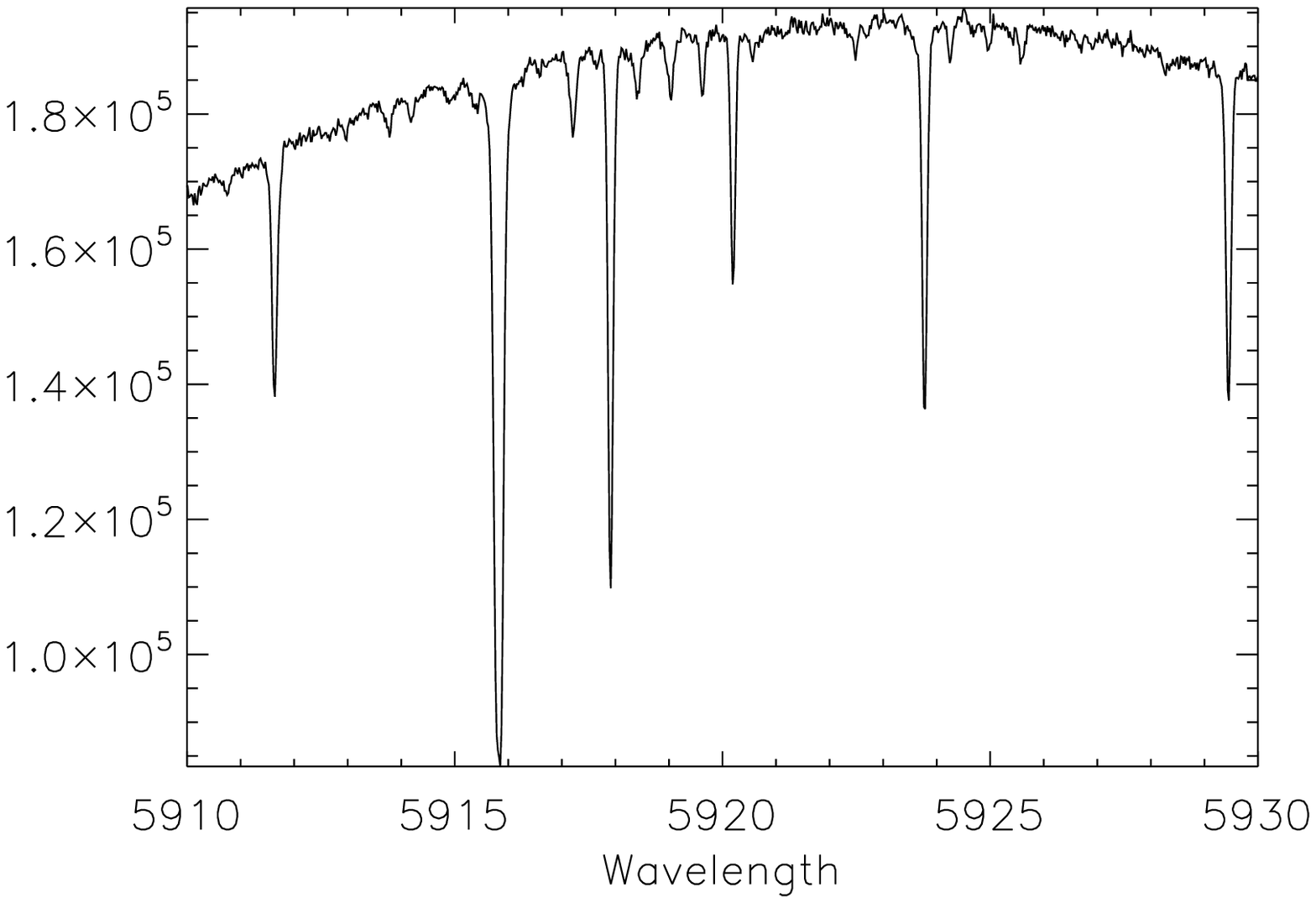}
\caption{\label{fig:spectrum}  Top: Fragment of the spectrum of a
nearby G8V star HD~20794 (82~Eri, $V=4.26$) in slicer mode recorded 
on  October 8 2011 with a 10-minute exposure. Bottom: portion of 
the extracted order with signal-to-noise ratio (SNR) of $\sim$400.}
\end{figure}

All CHIRON data  are processed at Yale and  distributed to other users
in   both    raw   and   extracted,    wavelength-calibrated   formats
(Fig.~\ref{fig:spectrum}).  Processing  is  done every  morning  after
observations have been acquired. A text file for each night containing
the  list   of  files  and  their  relevant   parameters  is  produced
automatically and used to organize the processing.

First, raw images are corrected for  bias and gain in each of the four
amplifiers  to get  the combined  image in  electrons per  pixel.  The
quartz   exposures    are   median-averaged   for    flat-field   (FF)
calibration. In the  fiber-fed CHIRON, we do not  have the possibility
to project a wide quartz spectrum for pixel-by-pixel FF correction, as
in  classical  slit  spectrographs.    Instead,  the  FF  spectrum  is
normalized by a smoothing polynomial  in each order and the flattening
calibration is done by dividing the extracted object spectra by the FF
spectrum.   This method  corrects sensitivity  variations of  the CCD,
provided  that  both  star  and  quartz  spectra  are  extracted  from
identical pixels. The scattered  light between the orders is evaluated
and subtracted. 

Stability of the order position on the CCD is key to spectrum extraction. 
The polynomial coefficients describing orders are defined from the quartz 
exposures and applied to the stellar spectra. The slit modes differ only by 
vertical shift and width of the orders, so the same set of order coefficients 
(refreshed nightly) serves for all modes. This is particularly useful for the 
slicer mode where the inter-order spacing at long wavelengths is less than 
the order width. Our standard processing algorithm delivers a set of 62 fixed orders 
covering the range of 450--885\,nm. It discards shorter wavelengths and edges of the orders.

The spectrum  extraction algorithm is  based on the REDUCE  package by
\citet{REDUCE}.   The spectrum profile  across the  order is  used for
removal    of   cosmic-ray   spikes    and   for    optimum   spectrum
extraction. However, we prefer  the boxcar extraction with fixed order
width, rather  than the  optimum extraction.  This  choice is  made to
ensure that  the flat-field correction  uses exactly same  pixels with
the same (unit)  weight. We checked that for faint  stars the noise in
box-car  extraction  is  insignificantly  higher  than  the  noise  in
optimally extracted spectra.

Wavelength calibration  is done by matching  nightly extracted spectra
of a ThAr lamp to a cataloged list of lines. This is done automatically,
using previous solutions as an  initial guess. The reduced spectra are
written in  the FITS files  with three dimensional arrays.   The first
plane of the array contains the wavelength for each pixel and for each
order number and is assigned  from the nearest (in time) ThAr exposure
in the  same slit  mode. The  second plane of  the array  contains the
extracted intensity (in electrons) for each pixel and order.

\subsection{Precise RVs with iodine reference}
\label{sec:iodine}

The program observations with CHIRON are taken with an iodine cell and
we employ a forward  modeling technique \citep{Butler96} to derive the
Doppler shifts.   The modeling process requires  an intrinsic spectrum
of both  the target  star and the  iodine cell.  These  are multiplied
together, with a Doppler shift as a free parameter, and convolved with
the SLSF to model our program observations taken with the iodine cell.

The intrinsic spectrum  of the iodine cell is  obtained by the Fourier
Transform Spectrometer (FTS) scan of our iodine cell with $R=800\,000$
and SNR~$\sim $1000.   To derive the intrinsic spectrum  of our target
stars, we obtain a one-time set of three or more $R=136\,000$ template
observations taken without the iodine  cell and co-add these spectra to
clean  cosmic rays  and  to build  up  the SNR.  The co-added  template
observation must  then be de-convolved  to yield the  intrinsic stellar
spectrum that  is used in  our Doppler analysis. The  deconvolution is
carried  out using  a featureless  B-star as  a light  source  for the
iodine cell; we use a Levenberg-Marquardt algorithm to derive both the
iodine wavelength solution  and the SLSF that degrades  the FTS iodine
spectrum to provide a best fit for our B-star iodine observation. This
SLSF is  adopted for the  adjacent template deconvolution  to generate
the intrinsic stellar spectrum.  The B-star iodine observation is also
used  to  assign  a  wavelength  solution  to  the  intrinsic  stellar
spectrum.

An important  advantage of the  CHIRON fiber-fed spectrometer  is that
the SLSF  is much less variable  than for a  slit-fed spectrometer. In
tests at Keck  HIRES, the SLSF was  found to be a factor  of ten times
more  stable  with  standard  circular  fibers compared  to  slit  fed
spectrometers  \citep{Spronck13b}. Additional  tests in  the  lab show
that the  SLSF stability is further  improved with the  use of optical
fibers that  have an octagonal  cross-section \citep{Spronck12a}. This
improved  modal  scrambling occurs  because  circular  fibers tend  to
preserve  the  angle of  incidence  throughout  the  wave guide  while
octagonal fibers  break the rotational symmetry. 
This  coupling  stability   from  the  fiber  is  particularly
important for  the template deconvolution  described above and  it can
also be used  to constrain the range of SLSF  variability in models of
our program observations.

When constructing a Doppler model of wavelength shifts for our program
observations, the spectrum is broken into smaller wavelength segments,
or  ``chunks''  along  the  iodine  orders.  Each  chunk  is  analyzed
independently so that  the spatially varying SLSF can  be modeled over
the two dimensional  CCD format. For CHIRON, we  analyze 760 chunks in
twenty orders and each chunk gives a nearly independent measurement of
the  stellar velocity.   The measurement  is not  entirely independent
because  we carry  out local  averaging  of the  SLSF for  neighboring
chunks in the last pass of the Doppler code. The standard deviation of
the 760 chunk velocities yields the formal measurement uncertainty for
each observation.  This measurement uncertainty  is a function  of the
SNR.   When several  observations  are  taken in  a  given night,  the
velocities  can be  averaged  and  the nightly  error  bars will  then
decrease   with   the   square   root   of  the   number   of   binned
observations. The nightly errors are generally smaller than the RV rms
scatter over several  nights. The RV can change  over longer time-lines
because  of systematic  errors in  the analysis,  instrumental errors,
dynamical velocities  in the star (e.g. from  planetary companions) or
from  photospheric signals  that introduce  spectral  line asymmetries
such as star spots or  variability in granulation. We adopt the binned
nightly  velocity  uncertainties  as  a measure  of  our  instrumental
precision and the velocity rms over several nights as a measure of our
instrumental stability.

\begin{figure}[ht]
\epsscale{1}
\plotone{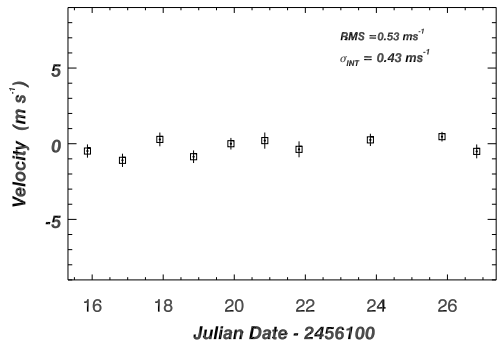}
\caption{\label{fig:rv_tauceti} Observations of HD~10700 ($\tau$ Ceti) taken  
beginning July 6, 2012 after upgrading CHIRON yield a nightly 
RV precision of 0.43\,\ms ~and the rms scatter over 12 nights is 0.53\,\ms.}
\end{figure}

In  Figure  ~\ref{fig:rv_tauceti},  we  show  a  12-nights  set  of  RV
measurements for the chromospherically inactive star, HD~10700 ($\tau$
Ceti), obtained after upgrading CHIRON.  The average SNR is 140 in the
individual observations.  This SNR is controlled by the exposure meter
and  our  Doppler  analysis  yields single  measurement  precision  of
0.82\,\ms.  We  took six  consecutive observations and  averaged these
velocities to obtain a binned  nightly precision of 0.43\,\ms. The rms
of  this  12-nights  data  set  is  0.53\,\ms.  To  the  best  of  our
knowledge, this  is both the  highest precision and RV  stability that
has been achieved with the iodine Doppler analysis technique.

\section{CHIRON performance}
\label{sec:perf}

\subsection{Efficiency and limiting magnitude}

\begin{figure}[ht]
\epsscale{1.0}
\plotone{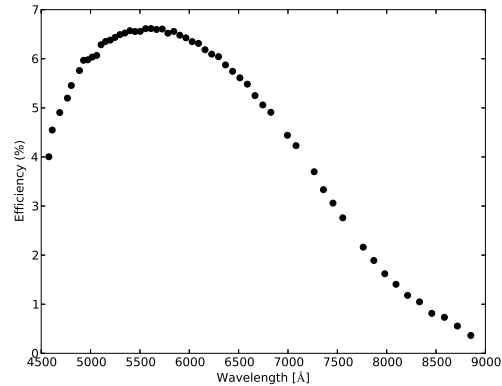}
\caption{\label{fig:eff} Spectral  efficiency of CHIRON  in fiber mode
  measured on  August 1, 2013  (median for three standard  stars, with
  three  exposures  of  each  star). The  atmospheric  extinction  was
  removed, so  that the  curve includes only  telescope, spectrograph,
  and detector.  }
\end{figure}

The internal transmission of CHIRON  is very high, better than 0.50 in
all modes.   The octagonal  fiber feed is  also quite  efficient, with
transmission of  0.87 at 514\,nm (including  FRD losses).  Considering
the known reflectivity of the two telescope mirrors, light loss at the
fiber  aperture due  to  seeing,  other optics,  and  the CCD  quantum
efficiency, we expected the total  efficiency at 500\,nm (ratio of the
photon flux outside the atmosphere to the signal in electrons at blaze
peak) to be at least  0.12.  Yet, direct measurement with three A-type
stars  performed on  June 5,  2012 (after  the upgrade)  indicate peak
efficiency of only  6\%.  The measurements repeated  on August 1,
2013 using another three standards and reduced by a slightly different
method,  gave essentially the  same result  (Fig.~\ref{fig:eff}). We
attribute poor efficiency at wavelengths beyond 600\,nm to the reduced
transmission of the octagonal fiber which was optimized for the iodine
region and  not characterized  at longer wavelengths.   The efficiency
can still  be affected by  guiding and/or focusing.  Indeed,  the flux
from $\alpha$~Cen  on some  nights is at  least two times  larger than
average and agrees well with the expected total efficiency.

\begin{figure}[ht]
\epsscale{1.0}
\plotone{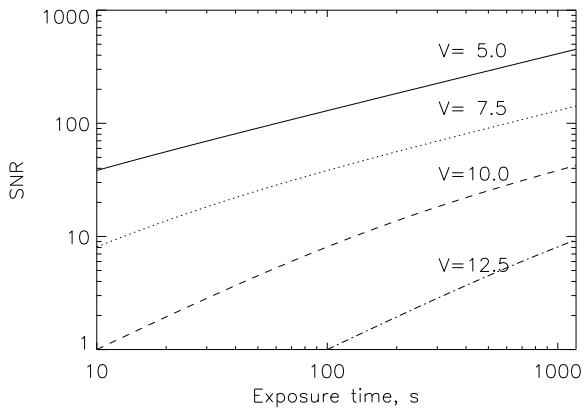}
\caption{\label{fig:etc} Signal-to-noise  ratio per pixel  vs. stellar
  $V$-band magnitude and exposure time  in the slicer mode (at 550\,nm near
  the blaze peak).}
\end{figure}

Calculation of the signal-to-noise ratio per pixel is based on the following formula:
\begin{equation}
{\rm SNR} = N_{ph}/\sqrt{ N_{ph} + K R^2} ,
\label{eq:SNR}
\end{equation}
where $N_{ph}$  is the  number of stellar  photons per  spectral pixel
collected during the  exposure time, $R=5.5$ is the  CCD readout noise
in electrons, $K$ is the number  of binned pixels across the order. In
the 3$\times$1 slicer  mode $K=9$, the pixel size  is about 0.0197\AA~
at  a  wavelength of  550\,nm.   In  the  fiber mode  with  4$\times$4
binning, $K=2.5$ (assuming optimum  extraction) and the spectral pixel
is 4  times larger.  We calculate  $N_{ph}$ by adopting  the 6\% total
efficiency and  obtain plots like  one in Fig.~\ref{fig:etc}.   In the
regular or narrow slit modes,  the efficiency is reduced by additional
slit losses.  In the fiber mode,  the sensitivity is higher and we can
reach SNR=20 on a $V=12.5^m$ star  in a 15-min exposure.  Noise in the
extracted featureless  spectra is indeed  in agreement with  the above
SNR estimates.

\subsection{Stability}

The effort  to stabilize the  instrument resulted in the  increased RV
precision  with  iodine  calibration  (\S~\ref{sec:iodine}).   The  RV
stability with the standard  ThAr wavelength calibration taken immediately
before  or after  program star  and cross-correlation  of  the stellar
spectrum  was  shown to  be  about 7\,\ms  ~on  the  35-day time  span
\citep{Tokovinin};   possibly it can  be improved further.   This opens
interesting applications  to stars that  are too faint for  the iodine
technique.

\subsection{Spectrum quality}

On bright  stars, very high SNR  can be reached  (e.g. SNR$\sim$400 in
Fig.~\ref{fig:spectrum}). The  ultimate limit of the SNR  in CHIRON is
yet to  be established.   During CHIRON integration,  we used  a green
laser  and were  surprised by  the lack  of parasitic  reflections and
ghosts: all light went into one location on the CCD. The simplicity of
the optical  scheme and good  coatings contribute to the  cleanness of
the resulting spectra.

\section{Summary}
\label{sec:concl}

CHIRON   is  a   high  resolution   (27\,000  to   136\,000)  facility
spectrometer  on the 1.5-m  telescope at  CTIO.  The  spectrometer was
built  with   supplemental  funding  to  the   American  Recovery  and
Reinvestment Act through  the NSF MRI program in  2009. The instrument
was  commissioned in  March 2011  and  is available  to the  community
through      telescope       time      partnerships      with      the
SMARTS\footnote{http://www.astro.yale.edu/smarts/} or through the NOAO
time allocation  committee.  Observing  is executed by  CTIO telescope
operators and  a data reduction pipeline for  supported modes provides
extracted spectra for SMARTS  observers the day after observations are
obtained.

In  January 2012,  we  carried  out several  upgrades  to improve  the
throughput and  stability of CHIRON  for the purpose of  improving the
precision of RV measurements.  The upgrade effort included an exposure
meter to  calculate the photon-weighted midpoint  for observations and
to auto-terminate observations at a user-specified SNR, replacement of
the echelle grating with one  of a higher efficiency, vacuum enclosure
for  the echelle grating  to minimize  variability in  the dispersion,
replacement  of the  optical fiber  feed with  an octagonal  fiber, AR
coating  (sol-gel) for  optical surfaces,  development of  a reduction
pipeline,  program scheduling and  quality control  software, improved
thermal  control of  the instrument,  and  a new  CCD controller.   To
assess the  impact of these  hardware upgrades to CHIRON,  we compared
our  current RV precision  for HD~10700  (Figure \ref{fig:rv_tauceti})
with measurements obtained in  2011 (before upgrading CHIRON). This is
not a perfect comparison because  we cannot control for differences in
stellar  activity  or  potential  dynamical velocities  from  planets.
However, we see a significant improvement; even with a slightly better
average SNR of 155,  the typical single-measurement uncertainty before
the upgrade was 1.5\,\ms\ and the lowest rms over any two week stretch
in 2011 was 1.91\,\ms.  Based  on this comparison, we conclude that we
gained more than a factor of two improvement in the single measurement
precision.   We  also  controlled  systematic  errors,  improving  the
instrumental  stability,   as  evidenced  by  the   factor  of  $\sim$4
improvement in the velocity rms over several nights.

To  date, four publications  that have  made use  of CHIRON  data have
appeared             in             peer-reviewed             journals
\citep{Tokovinin,Shore,Ingleby,Ratajczak},   with   additional  papers
submitted and in preparation.

\acknowledgments
The CHIRON was designed and fabricated at the CTIO workshop. 
DAF acknowledges support for construction of this instrument through 
the NSF MRI grant 0923441. DAF further acknowledges support for 
instrument upgrades, software development, and a precision Doppler planet 
search from Yale University, NSF AST-1109727, and NASA NNS12AC01G.  
DAF thanks contributors to The Planetary Society for support 
that advanced our understanding of fiber coupling of spectrographs and provided 
telescope time.



\end{document}